\documentclass{appolb}
\usepackage{epsfig}
\usepackage{mcite}
\usepackage{wrapfig,rotating}
\newcommand{\kt}{k_{t}}
\def\CASCADE{{\sc Cascade}}
\newcommand{\pom}{I\!\!P}

\begin{document}
\title{Vector meson cross sections at HERA%
\thanks{Presented at ISMD 07}%
}
\author{Hannes Jung \\ 
{\it on behalf of the H1 and ZEUS Collaborations}
\address{DESY, Hamburg, FRG}
}
\maketitle
\begin{abstract}
Inelastic and elastic (exclusive) cross section measurements of vector meson production at
HERA are discussed.
\end{abstract}
  
\section{Introduction}
Vector meson production is an ideal tool for studying the structure of the
proton and to investigate the transition from purely soft to hard pQCD
processes. 
\begin{wrapfigure}{l}{0.25\columnwidth}
\begin{center}
\vspace*{-0.8cm}
\includegraphics[width=0.15\textwidth]{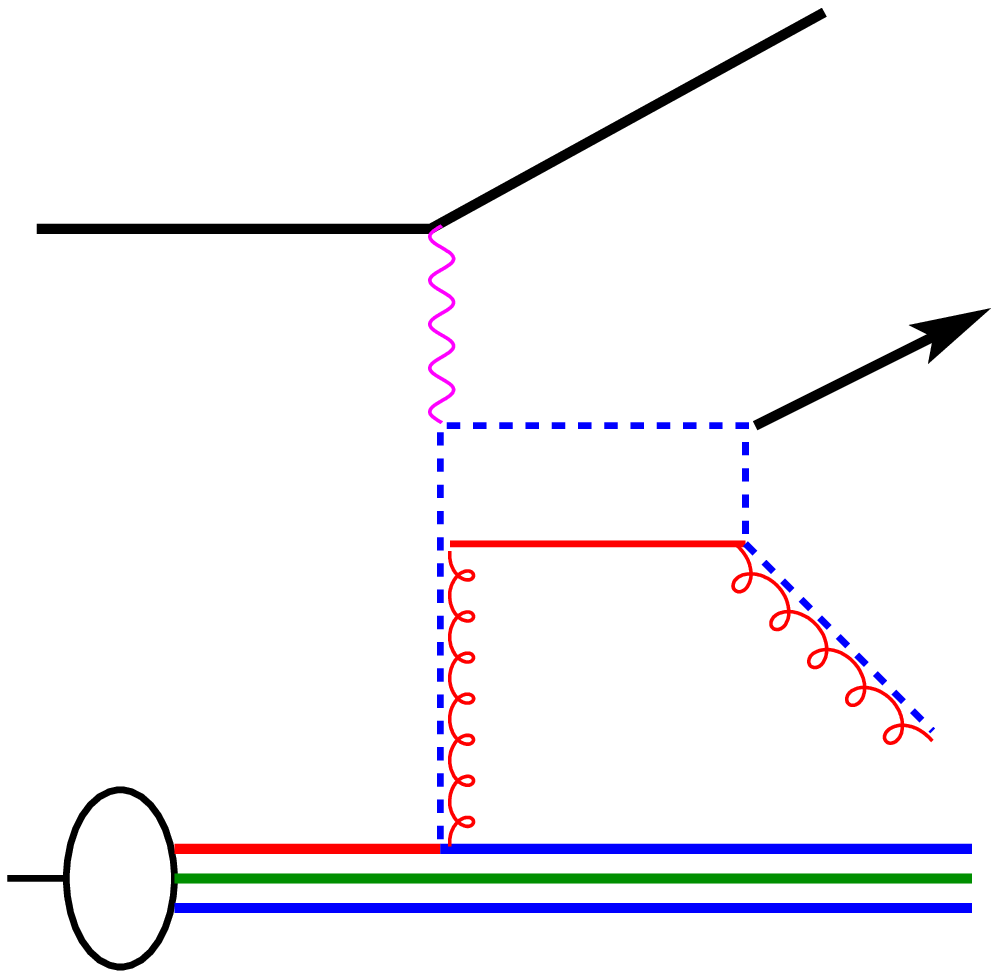}
\vspace*{-2cm}
$(a)$\\
\vspace*{2.0cm}
\includegraphics[width=0.15\textwidth]{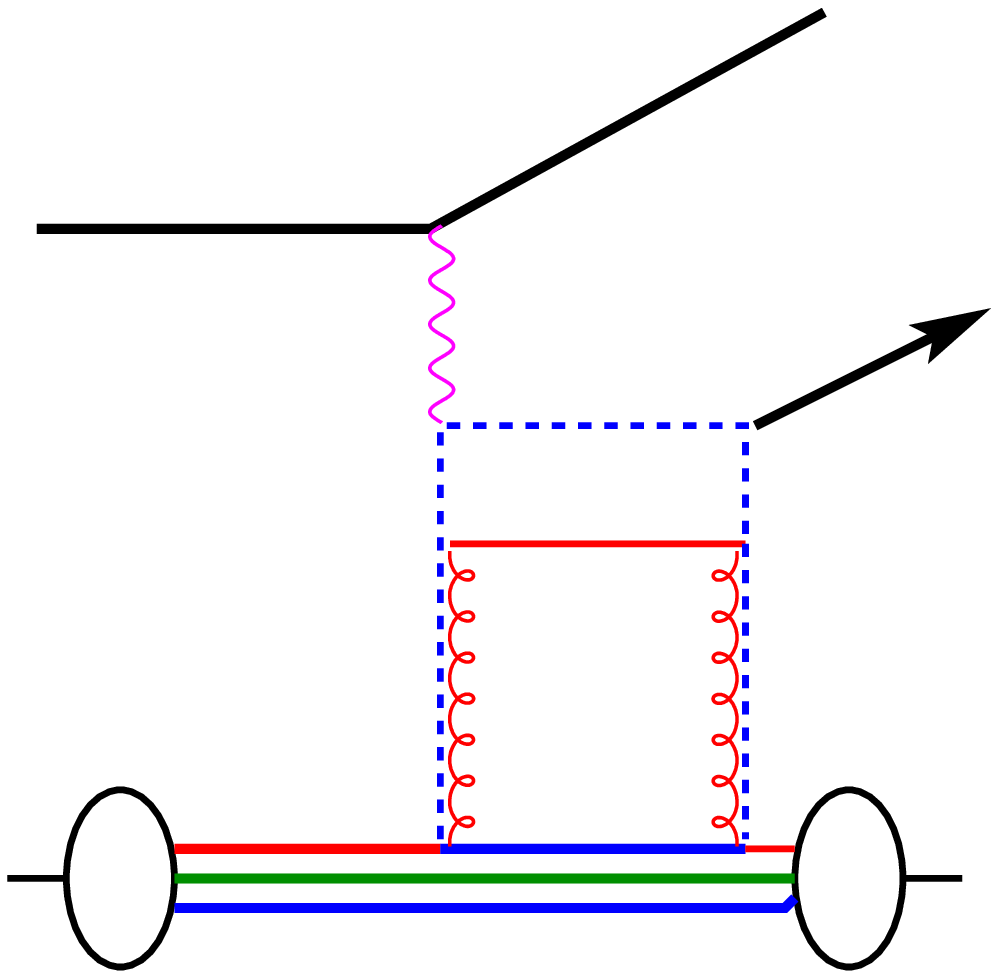}
\vspace*{-1cm}
$(b)$
\vspace*{1.0cm}
\caption{\label{vm-prod}
Schematic picture of inelastic $(a)$ and elastic $(b)$ vector meson lepto-production}
\end{center}
\vspace*{-1.3cm}
\end{wrapfigure}
Vector mesons can be produced in two different
ways in lepton proton scattering, in so-called inelastic processes, where
the proton breaks, or elastic (exclusive) processes, where the incoming proton
stays intact. The different production mechanisms are shown
schematically in Fig.~\ref{vm-prod}. They suggest that the cross section for
 inelastic vector meson production behaves like 
 $\sigma_{inel} \sim xG(x,\mu^2)$ whereas
for elastic production like  
$\sigma_{el} \sim \left[xG(x,\mu^2)\right]^2$. Different regions of the
available phase space ranging from photoproduction
($Q^2\sim 0$) to the DIS
regime ($Q^2 > 1$~GeV$^2$) and various vector mesons,
including photons can be investigated.
\vspace*{-0.4cm}
\section{Inelastic Vector Meson Production}
Inelastic $J/\psi$ production (see Fig.~\ref{vm-prod}$(a)$) has been measured by
the H1 and ZEUS experiments both in the 
photoproduction ~\cite{H1prelim-07-172,Chekanov:2002at} and
DIS~\cite{H1prelim-07-172,Chekanov:2005cf} regime.  
\begin{figure}[h]\vspace*{-0.6cm}
\begin{minipage}{0.46\columnwidth}
\includegraphics[width=1.0\textwidth]{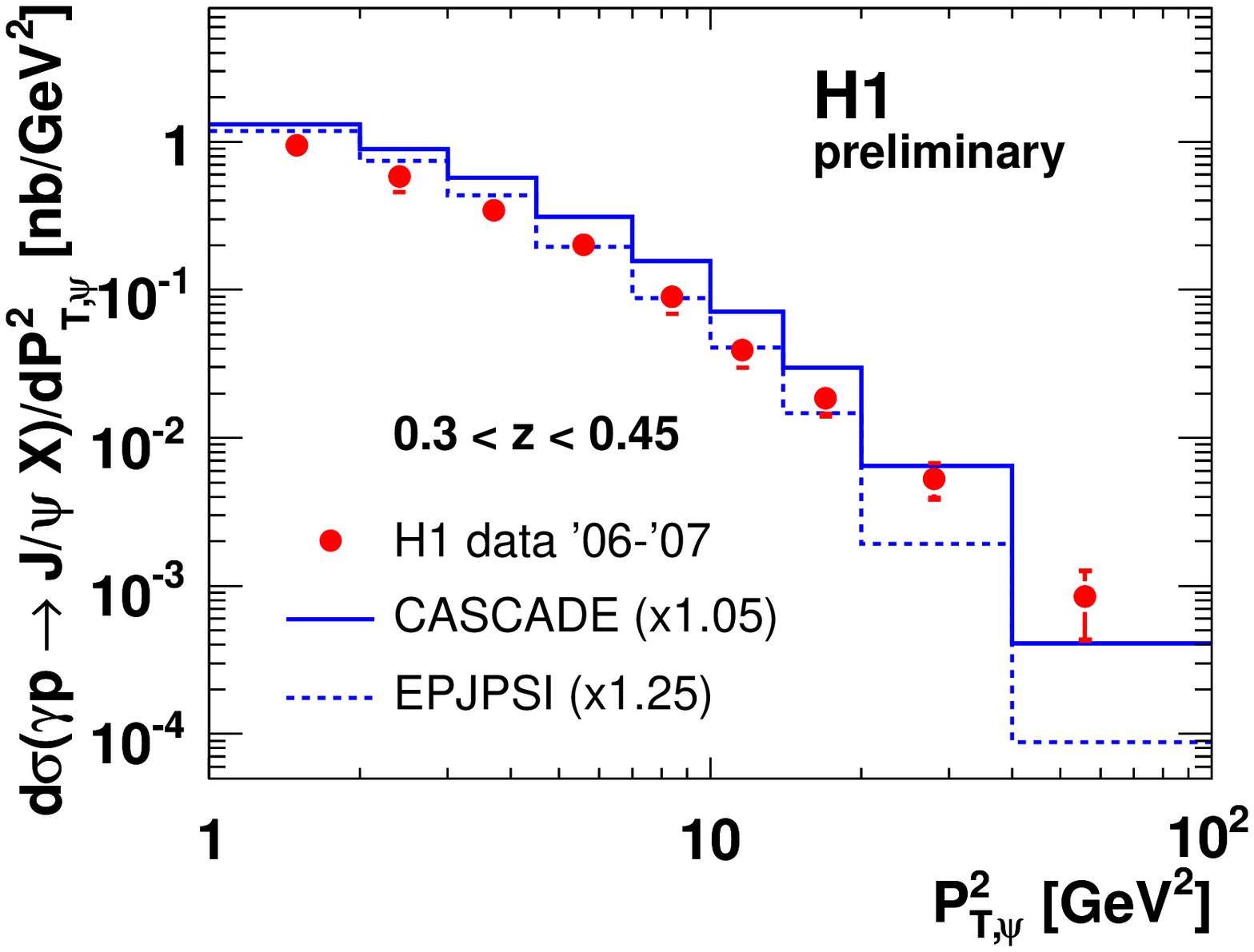}
\caption{\label{jpsi-inel-ph} Cross section as a function of $p_t^2$ in
photoproduction of $J/\psi$ mesons \protect\cite{H1prelim-07-172}}
\end{minipage}\hspace*{0.5cm}
\begin{minipage}{0.46\columnwidth}
\includegraphics[width=1.0\textwidth]{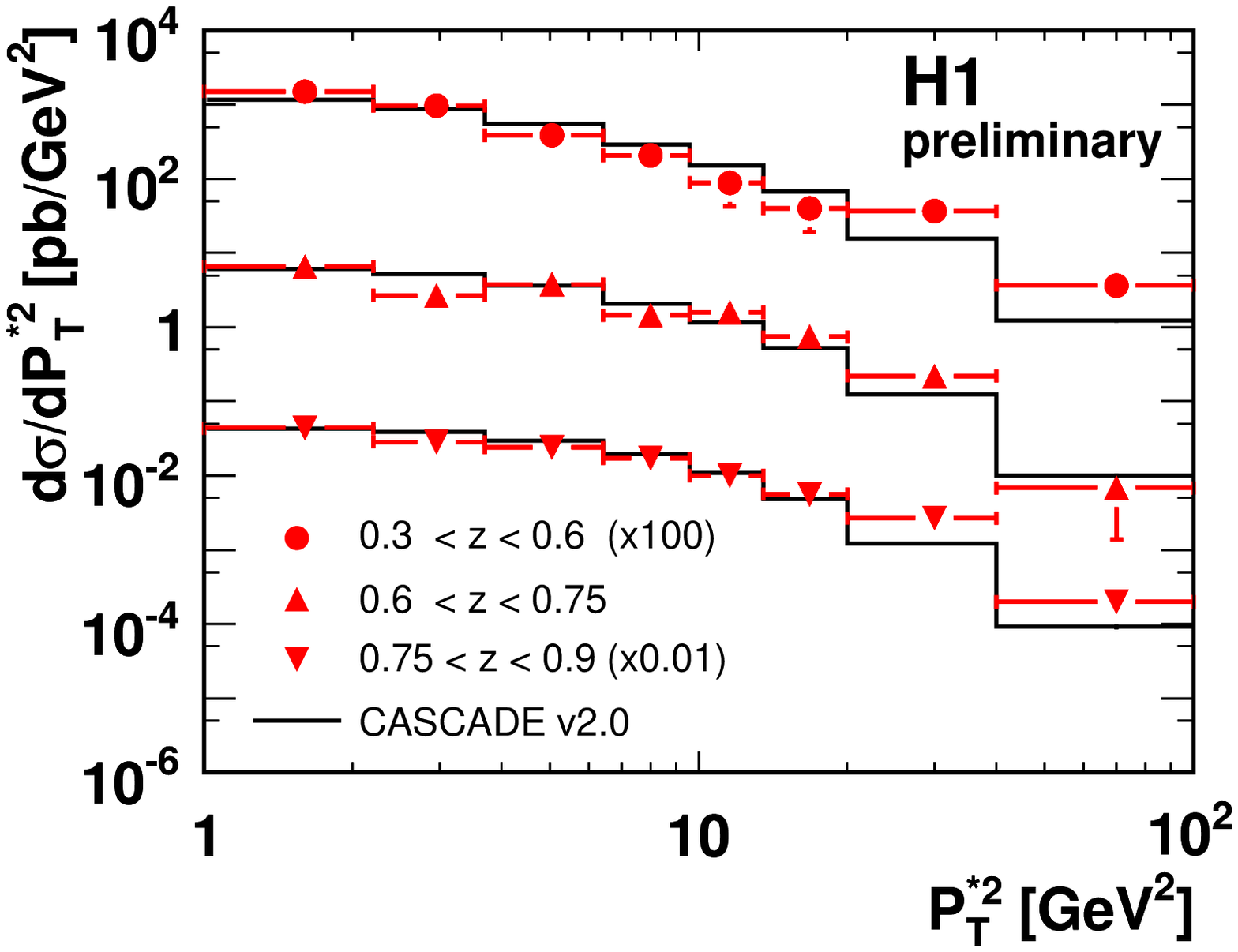}
\caption{\label{jpsi-inel-dis}
Cross section as a function of $p_t^2$ for $J/\psi$ lepto-production
in the range of $3.6 < Q^2 < 100 $~GeV$^2$\protect\cite{H1prelim-07-172}
}
\end{minipage}\vspace*{-0.1cm}
\end{figure}
The cross section $\frac{d\sigma}{dp_t^2}$ as measured by H1\cite{H1prelim-07-172} for 
photoproduction is shown in Fig.~\ref{jpsi-inel-ph}. 
In Fig.~\ref{jpsi-inel-dis} the cross section $\frac{d\sigma}{dp_t^2}$ in the
DIS region~\cite{H1prelim-07-172} ($3.6 < Q^2 < 100 $~GeV$^2$) for
different values of the inelasticity 
$z=\frac{p_p.p_{J/\psi}}{p_p.p_{\gamma}} =
\left. \frac{E_{J/\psi}}{E_\gamma} \right|_{\footnotesize\mbox{p rest}}$ is shown. 
The results are in good agreement with the measurements of 
ZEUS~\cite{Chekanov:2002at,Chekanov:2005cf} and they 
agree well in normalization and shape with a
QCD calculation~\cite{Lipatov:2002tc,*Baranov:2003at,*Baranov:2002cf} using
$\kt$-factorization (implemented in \CASCADE\ ~\cite{CASCADEMC}).
The NLO calculation of \cite{Kramer:1994zi} applicable for the photoproduction region
also agrees well with the measurement.
\par
Prompt photon production (the vector meson is replaced by a real photon
in Fig.~\ref{vm-prod}) in  $Q^2\sim 0$ and in the
DIS region have been measured by H1 and ZEUS
\cite{prompt-photon-h1,*prompt-photon-zeus,*promot-photon-dis-h1,*prompt-photon-dis-zeus}.
Calculations using leading log parton showers or NLO calculations (${\cal
O}(\alpha^3\alpha_s)$) are able to
describe some of the features of the data in DIS but not all. In the
photoproduction region the measurements are reasonably well described in shape
by NLO calculations but the predicted cross sections are a factor $\sim 2$
smaller than the measured ones, indicating that
significant contributions (presumably higher order corrections) are still missing in the
calculations. 
\par
Thus, vector meson production (not prompt photon production) is reasonably well understood and
can be used to further investigate the structure of the hadronic final state: 
the elastic (exclusive)
production of vector mesons, where the proton stays intact.  
\vspace*{-0.4cm}
\section{Elastic (exclusive) Vector Meson Production}
\begin{figure}[h]\vspace*{-0.5cm}
\begin{minipage}{0.48\columnwidth}
\includegraphics[width=0.7\textwidth]{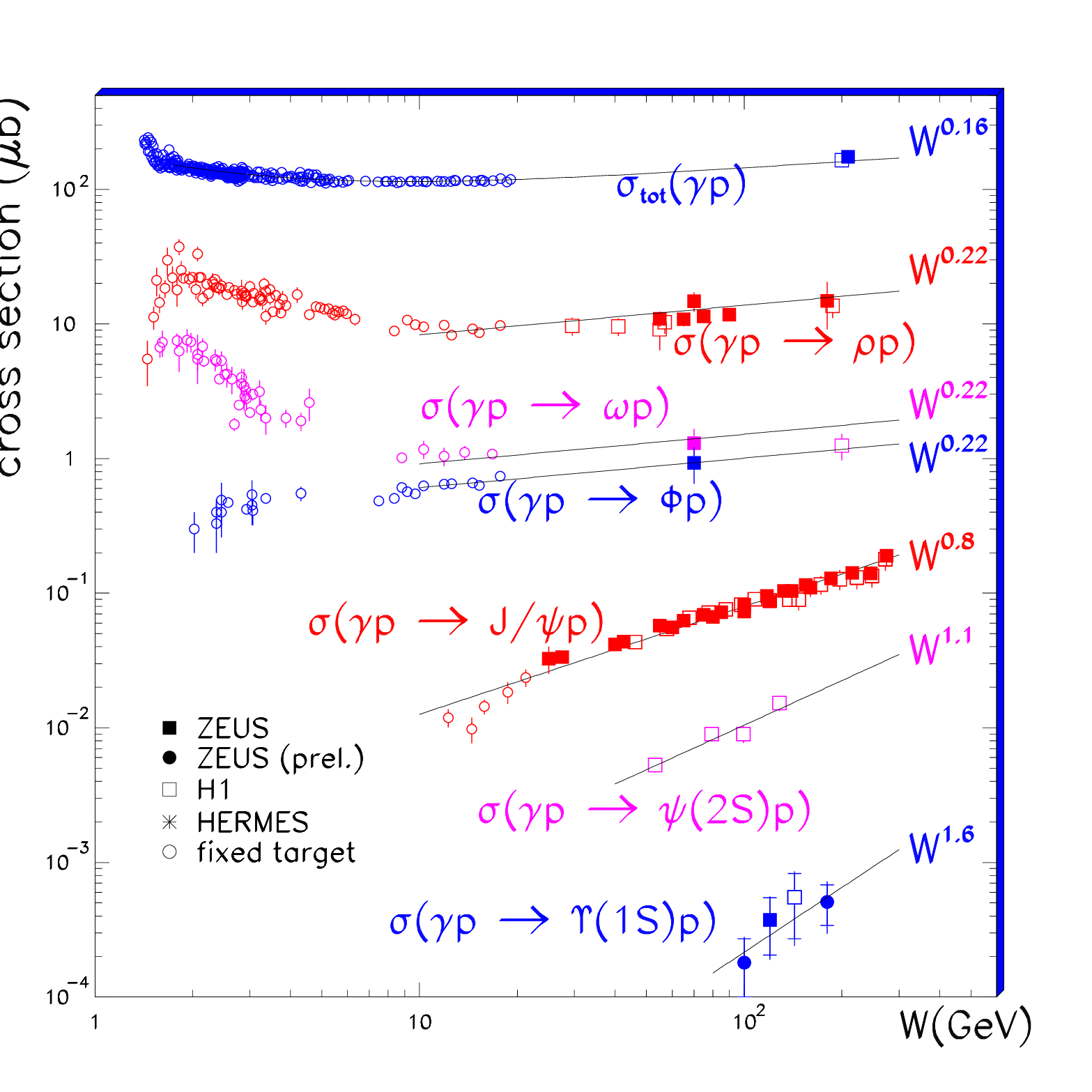}
\caption{\label{excluisve-vm-prod-W}
Cross section for $\gamma^*p \to Vp$ as a function of $W$ The total $\gamma p$
cross section is also shown. The lines are results of a fit of the form
$W^{\delta}$~\protect\cite{ZEUS-prel-07-015}.
}
\end{minipage}\hspace*{0.5cm}
\begin{minipage}{0.48\columnwidth}
\vspace*{-0.4cm}
\includegraphics[width=0.7\textwidth]{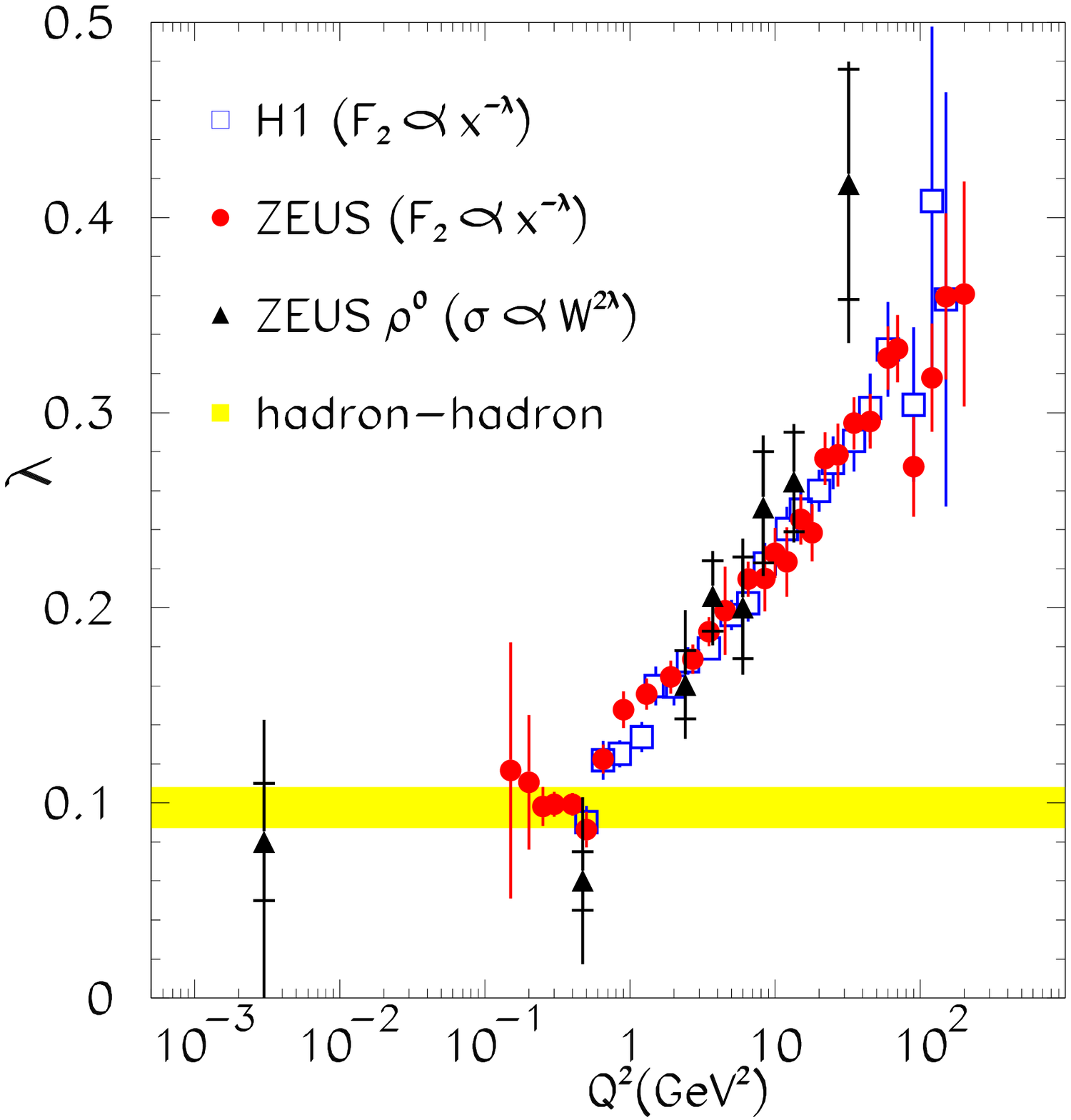}
\caption{\label{vm-prod-lambda}
$\lambda$ for $\rho^0$ production at large $Q^2$ compared with
the one from inclusive $F_2$~\protect\cite{Levy}}
\end{minipage}\vspace*{-0.5cm}
\end{figure}

The cross section for elastic vector meson production as a function of the
$\gamma^*p$ center of mass energy $W$ is shown in
fig.~\ref{excluisve-vm-prod-W}~\cite{ZEUS-prel-07-015}.
One observes a steep rise of the cross section with $W$ for heavy vector mesons.
A similar behavior is observed for light vector meson production at large $Q^2$.
As suggested in Fig.~\ref{vm-prod}, elastic vector meson production at large
scales should behave like $\left[xg(x,\mu^2)\right]^2 \sim x^{-2 \lambda}$, whereas inelastic
processes, like inelastic vector meson production or the inclusive cross section
$F_2$ at small $x \sim 1/W^2$, behave like  $xg(x,\mu^2) \sim x^{\lambda}$. 
Fig.~\ref{vm-prod-lambda}~\cite{Levy} shows a measurement of $\lambda$ for $\rho^0$ production 
at large $Q^2$ and
compared with the one from $F_2$ showing a similar
energy dependence $\lambda$, in contrast to the naive expectation coming from
2-gluon exchange. Note, a similar behavior is seen in the energy
dependence of diffractive and inclusive cross sections at large
$Q^2$~\cite{Chekanov:2004hy,*Aktas:2006hy}.

\begin{figure}[h]\vspace*{-0.5cm}
\begin{minipage}{0.48\columnwidth}
\includegraphics[width=0.85\textwidth]{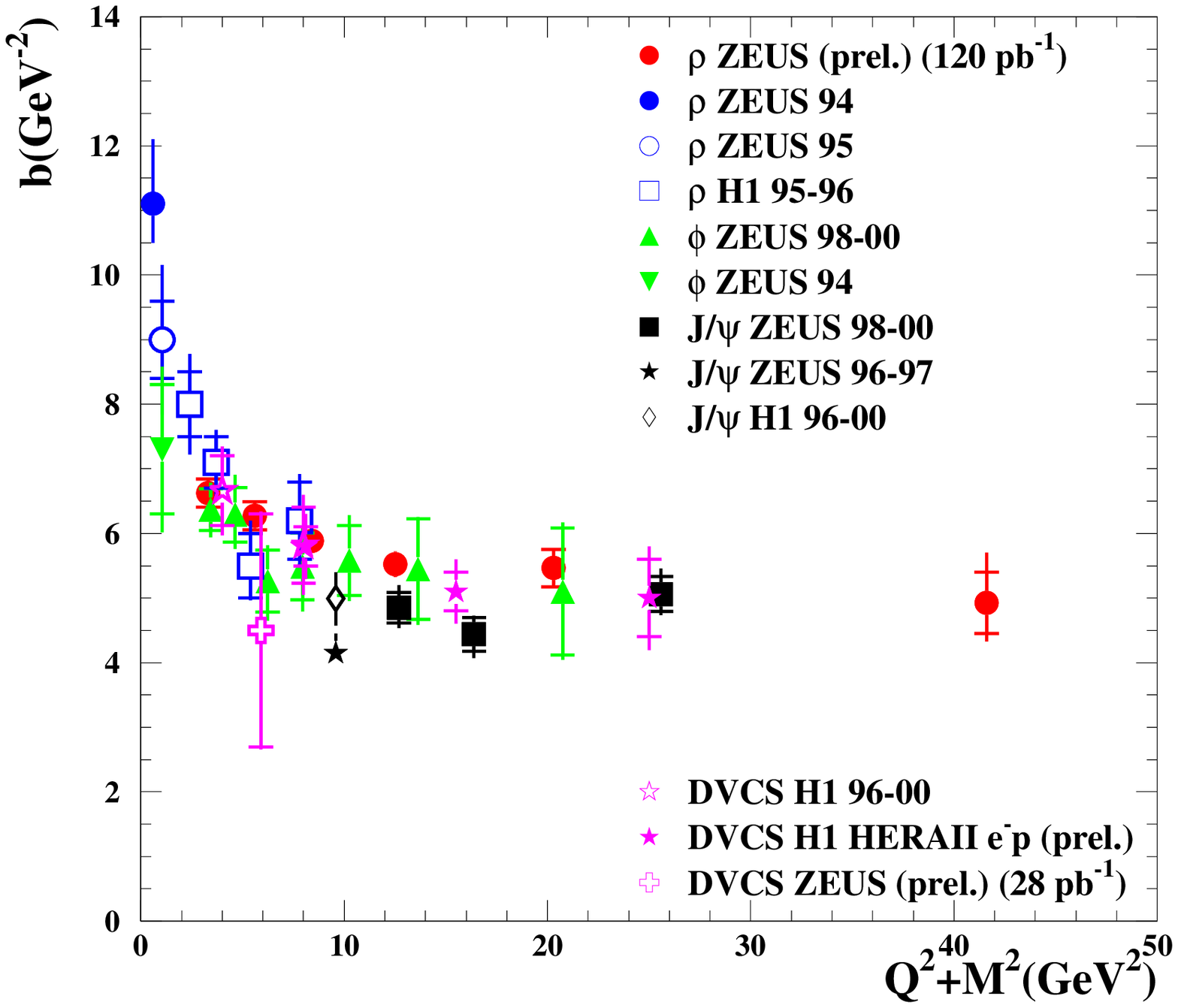}
\caption{\label{b-slope}
$b$-slope as a function of $Q^2+M^2$ for different vector mesons}
\end{minipage}\hspace*{0.5cm}
\begin{minipage}{0.48\columnwidth}
\includegraphics[width=0.7\textwidth]{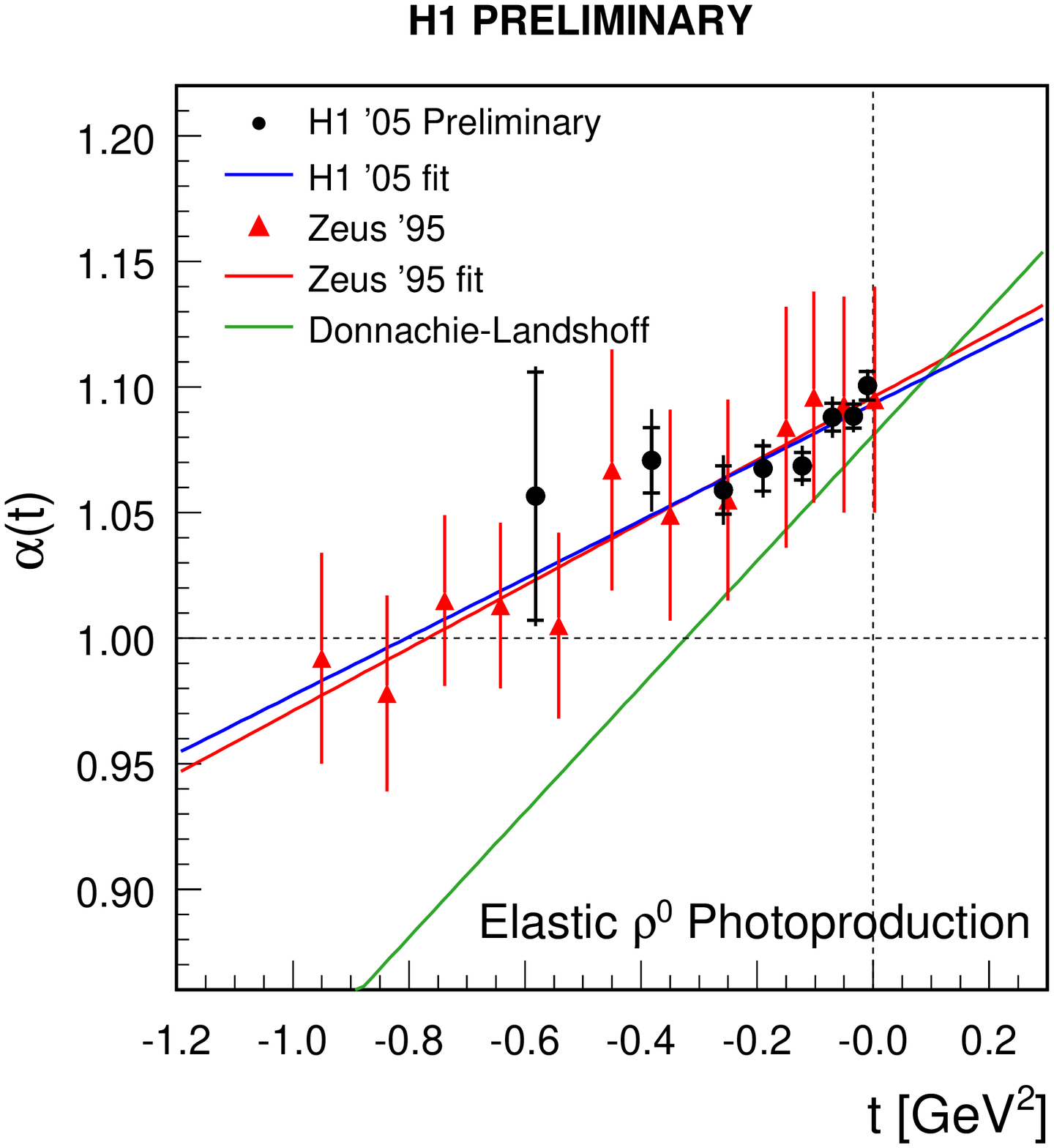}
\caption{\label{alpha_pom}
$\alpha_{\pom}(t)$ measured in elastic $\rho^0$ photoproduction. 
}
\end{minipage}\vspace*{-0.3cm}
\end{figure}
The dependence of the elastic vector meson production cross section on $t=(p -
p')^2$, the momentum transfer at the proton vertex, is often parameterized as
$\sigma \sim \exp{(- b |t|)}$. The measured $b$-slope obtained from different measurements
 of vector meson production is shown in 
 Fig.~\ref{b-slope}~\cite{ZEUS-prel-07-014} as a function of $Q^2 +
M^2$, with $M$ being the mass of the produced vector meson. At large $Q^2 +
M^2 \simeq 10$~GeV$^2$ the $b$-slope becomes constant at $b \sim 5 $~GeV$^{-2}$. 

The energy dependence of the cross section as a function of $t$ can be
parameterized with
$\frac{ d\sigma_{\gamma p}(W)}{dt}=
\frac{d\sigma_{\gamma p}(W_0)}{dt}\left( \frac{W}{W_0} 
\right) ^{4(\alpha_{\pom} (t) - 1) }$. 
The measurement of
 $\alpha_{\pom}(t)$ within a one experiment is shown in 
 Fig.~\ref{alpha_pom}~\cite{H1prelim-06-011} 
 and compared with results of~\cite{Breitweg:1999jy}. 
 Even in the photoproduction region the value
 of  $\alpha_{\pom}(t)$ is smaller than expected from soft hadron-hadron
 interactions.
\vspace*{-0.5cm}
\section{Conclusion}
Measurements of inelastic vector meson production in the photoproduction and
DIS region can be reasonably well described with higher order QCD calculations.
Measurements of elastic $\rho^0$ vector meson production show a energy dependence,
which is similar to the one obtained from inclusive measurements. Even in
photoproduction of elastic $\rho^0$ mesons, the measured $\alpha_{\pom}$ is
smaller than expected from soft hadron-hadron interactions. Thus understanding of
elastic vector meson production is still a challenge.
\vspace*{-0.8cm}
{\bibliographystyle{mybib} 
\raggedright 
\bibliography{ref}
}
\end{document}